\documentstyle [12pt,a4]{article}

\newcommand{\br}[1]{|#1\rangle}
\newcommand{\be}{\begin{equation}}
\newcommand{\ee}{\end{equation}}
\newcommand{\bea}{\begin{eqnarray}}
\newcommand{\ena}{\end{eqnarray}}
\newcommand{\beas}{\begin{eqnarray*}}
\newcommand{\enas}{\end{eqnarray*}}

\newcommand{\hsl}{U_q\widehat{sl}_2}
\newcommand{\hg}{U_q\hat{g}}
\newcommand{\g}{U_qg}

\newcommand{\D}{\Delta}
\newcommand{\ot}{\otimes}
\newcommand{\up}{\uparrow}
\newcommand{\dow}{\downarrow}

\newcommand{\nn}{\nonumber}
\newcommand{\la}{\lambda}

\newcommand{\hot}{\hat{\otimes}}

\newcommand{\op}{\oplus}
\renewcommand{\H}{{\cal H}}
\newcommand{\R}{\hat{R}}
\newcommand{\Pro}{{\cal P}}
\newcommand{\ab}[1]{{\rm{\bf #1}}}
\renewcommand{\dim}{{\rm dim}}
\newcommand{\id}{{\rm id}}
\newcommand{\gr}[2]{ {( #1,#2)} }
\newcommand{\V}{{\bf V}}
\newcommand{\W}{{\bf W}}
\newcommand{\Na}{{\bf N}}
\newcommand{\re}[1]{(\ref{#1})}
\newcommand{\vac}{|vac\rangle}
\begin{document}

\title{
Family of  Affine Quantum Group Invariant Integrable Extensions
of Hubbard Hamiltonian
}

\author{A.Avakyan\thanks{e-mail:{\sl avakyan@lx2.yerphi.am}},
            T.Hakobyan\thanks{e-mail:{\sl hakob@lx2.yerphi.am}},
\\and\\
            A.Sedrakyan\thanks{e-mail:{\sl sedrak@alf.nbi.dk}}\\
{\sl Yerevan Physics Institute,}\\
{\sl Br.Alikhanian st.2, 375036, Yerevan, Armenia}}

\date{September 1996}

\maketitle

\begin{abstract}

We construct the family of spin chain Hamiltonians, which have affine
quantum group symmetry $U_q\hat{g}$.
Their eigenvalues coincide with the
eigenvalues of the usual spin chain Hamiltonians,
but have the degeneracy of levels, corresponding
to affine $U_q \hat{g}$. The space of states of these spin chains is formed by
the tensor product of fully reducible representations of quantum
group.

The fermionic representations of constructed spin chain Hamiltonians
show that we have new extensions of Hubbard
Hamiltonians. All of them are integrable and have affine quantum group
symmetry.
The exact ground state of a such type model is presented, exhibiting
superconducting behavior via $\eta$-pairing mechanism.

\end{abstract}

\vfill
\rightline{YerPhI-1476(13)-96}

\newpage

\section{Introduction}

The interest to strongly interacting electron systems was raised after
discovery of high-$T_c$ superconductivity \cite{A}, where it is believed
that the interplay between magnetism and insulating behavior has a crucial
role. The Hubbard model \cite{H} is the simplest example of strongly
correlated electrons and was solved in
one dimension via
Bethe ansatz technique \cite{L}. Subsequently numerous publications appeared
 investigating one-band (multi-band) Hubbard models in one and two
 dimensions, most of which used approximate or numerical methods and only
few exact results are known. One of these exact results based on so called
$\eta$-pairing mechanism of superconductivity, the idea of which was
introduced
by Yang \cite{Yang,YZ} for the Hubbard model. It appeared, that there is
certain type of states exhibiting off-diagonal long-range order (ODLRO),
the concept, which had been developed in \cite{Y2}. In \cite{Y2,Sew,Ni}
was shown that ODLRO implies Meissner effect and flux quantization and
thus can be regarded as definition of superconductivity. But ground state
of simple Hubbard model is not superconducting \cite{Y2}. The first model
with truly superconducting ground state of $\eta$-pairing type is the
supersymmetric Hubbard model, introduced in \cite{EKS1,EKS2}. This
model is a certain extension of Hubbard model by additional nearest-neighbor
interactions and some restriction on hopping mechanism.

It appeared that $\eta$-pairing mechanism is not an exotic phenomenon and
recently some extended Hubbard models were defined, superconducting
ground state of which could be found exactly \cite{St,Ov,BKS,MC,Schad}.
The main property of these extensions is the fact, that they besides
the ordinary electron hopping and Hubbard interaction terms contain
also bond-charge interaction, pair-hopping, nearest-neighbor Coulomb
interaction and $XXX$ spin interaction terms.

In this article we develop the new technique and define a family of
other extensions of one dimensional Hubbard Hamiltonians, one
of which exhibits superconductivity based on
$\eta$-pairing. The defining property of this family is the fact, that they
have $U_q\widehat{sl}(2)$ affine quantum group symmetry.

Quantum group symmetry plays essential role in integrable statistical
models \cite{J85,J86_1,Dr86} and conformal field theory \cite{MR,PS,GS91}.

It is well known that many integrable Hamiltonians have quantum
group symmetry.
For example, $XXZ$ Heisenberg Hamiltonian with particular boundary terms
is $U_qsl_2$-invariant \cite{PS}.
The infinite $XXZ$ spin chain has larger symmetry:
affine $U_q\widehat{sl}_2$ \cite{DFJMN93}. The single spin site of
most considered Hamiltonians forms an irreducible representation of
Lie algebra or its quantum deformation.

In \cite{HS} we have constructed the family of spin chain Hamiltonians
${\cal H}$,
which have affine quantum group symmetry $\hg$, using  intertwining
operators between tensor products of
 its spectral parameter dependent reducible representations. These
intertwining operators commute with $\hg$ by definition, whereas
Hamiltonians ${\cal H}$
commute with $\hg$ by construction. The space of states of these spin chains
is formed by the tensor product of the fully reducible representations.
We have shown that the model, considered in \cite{RA}, which corresponds
to some extension of the Hubbard Hamiltonian in the strong repulsion limit,
is a particular case of our general construction. The affine quantum group
symmetry leads to a high degeneracy of energy levels.

The energy levels of these spin chains are formed on the states, constructed
from the highest weight vectors of quantum group representations. In particular
cases the restriction of the considered spin chains on these states gives
rise to Heisenberg model  or Haldane-Shastry long-range
interaction spin chain.
Hence, we have the generalizations of
these Hamiltonians, which have affine quantum group symmetry.

It appeared that  the fermionization of the simplest examples  of these
spin chain Hamiltonians
gives the family of extensions of Hubbard model.

In the sec.2 we give the definitions of quantum Kac-Moody group
$U_q\widehat{g}$ and properties of $R$-matrix.
In sec.3 we develop
the technique for construction of nearest-neighbor spin chain Hamiltonians,
which have
$U_q\widehat{sl}_2$ quantum group symmetry.
In sec.4 we perform this construction
to obtain Haldane-Shastry type long-range spin chain
Hamiltonians.

In sec.5 we consider the fermionization of
previously constructed integrable spin chains
for some simplest particular cases, in particular,
when the site is four dimensional and there is no dependence on
quantum group deformation parameter $q$.
We obtain in such way some integrable extensions of Hubbard model.
In the case, when the space of states of each site is a direct sum
of spin-$1$ and trivial multiplets, one obtain Hubbard model with
doping, pair hopping and nearest-neighbor interaction of holes.

In the case, when the space of states of each site is a direct sum
of two two dimensional representations, one obtain Hubbard model with
bond-charge interaction, density-density interaction and 'boson-boson'
interaction between nearest-neighbor hole and double occupied site.
We find the ground states in the fixed particle number sectors
and   phase space region, where this model exhibits $\eta$-pairing
superconductivity.
In the particular case, when the amplitude of last two interactions vanishes,
one obtain well known extended Hubbard model with bond-charge interaction
\cite{Klein,Schad}.

\section{Definitions}
\setcounter{equation}{0}

Let us recall the definition of quantum Kac-Moody group $\hg$
($\g$) \cite{J85,J86_1,Dr86}. It is
generated by elements $e_i, \ f_i, \ h_i$ satisfying the relations
\bea
\label{QG}
&{[}h_i,e_j{]}  = c_{ij}e_j \qquad {[}h_i,f_j{]}  = -c_{ij}f_j&
\\
& {[}e_i,f_j{]} = \delta_{ij}{[}h{]}_q &\nn
\ena
and $q$-deformed Serre relations, which we don't write.
Here $i=0,\dots,n$ for $\hg$ and $i=1,\dots,n$ for $\g$,
$q$ is a deformation parameter, ${[}x{]}_q:=
(x^q-x^{-q})/(q-q^{-1})$, $c_{ij}$ is a Cartan matrix of
corresponding affine Lie algebra $\hat{g}$ (finite Lie algebra $g$).

On $\hg$ ($\g$) there is a Hopf algebra structure:
\bea
\label{defcomul}
\D(e_i) =  k_i \ot e_i+e_i\ot k_i^{-1} &
\D(k_i^{\pm1}) = k_i^{\pm1} \ot k_i^{\pm1}\nn\\
\D(f_i) = k_i \ot f_i+f_i\ot k_i^{-1}
\ena
where $k_i:=q^{h_i\over 2}$.
This comultiplication can be extended to $L$-fold tensor
product by
\beas
\D^{L-1}(e_i) = \sum_{l=1}^L
	k_i \ot\ldots\ot k_i\ot\underbrace{ e_i}_l\ot k_i^{-1}
	\ot\ldots\ot k_i^{-1} \\
\D^{L-1}(f_i) = \sum_{l=1}^L
	k_i \ot\ldots\ot k_i\ot\underbrace{ f_i}_l\ot k_i^{-1}
	\ot\ldots\ot k_i^{-1} \\
\D^{L-1}(k_i^{\pm1}) = k_i^{\pm1} \ot\ldots\ot k_i^{\pm1}
\enas
There is an opposite comultiplication $\bar{\D}$, which is
obtained from \re{defcomul} by replacing $k_i^{\pm1}\to k_i^{\mp1}$.

For general $q$ the representations of quantum group $\g$ are is one to one
correspondence to the representations of nondeformed Lie algebra $g$.
Denote by $\V_\la$ the irreducible $\g$-multiplet with highest weight
$\la$. It possesses a highest weight vector $v^0_\la$, such that
\be
\label{hww}
e_i v^0_\la=0 \quad h_i v^0_\la=\la(h_i) v^0_\la, \quad i=1,\dots,n
\ee
Let $\hat{g}$ be an affine algebra and $g$ be its underlying finite
algebra. Then
for any complex $x$ there is the $q$-deformation of
loop homomorphism $\rho_x$:
$\hg\rightarrow\g$, which is given by \cite{J85}
\[
\begin{array}{lll}
\rho_x (e_0) = xf_\theta & \rho_x (f_0)= x^{-1}e_\theta &
	 \rho_x(h_0)=-h_\theta \\
\rho_x (e_i) = e_i & \rho_x (f_i) = f_i & \rho_x (h_i) = h_i,
\end{array}
\]
where $i=1\dots n$ and
$\theta$ is a maximal root of $\g$. Using  $\rho_x$ one can construct
the spectral parameter dependent representation  of $\hg$  from
the   representation   of $\g$.

Let $\V_1(x_1)$ and $\V_2(x_2)$ are constructed in such way irreducible finite
dimensional
representations of $\hg$ with parameters $x_1$ and $x_2$ correspondingly.
 The $\hg$-representations on $\V_1(x_1)\ot \V_2(x_2)$ constructed
by means of both $\D$ and $\bar{\D}$ are  irreducible, in general, and
equivalent:
\[
R(x_1,x_2)\D(g)=\bar{\D}(g)R(x_1,x_2), \quad \forall g\in \hg
\label{Rmatrix}
\]
The $R$-matrix $R(x_1,x_2)$ depends only on $x_1/ x_2$ and is a Boltsmann
weight of some integrable statistical mechanic system.

\setcounter{equation}{0}
\section{Quantum group invariant Hamiltonians for reducible
	representations}

Let $\V=\oplus_{i=1}^N \V_{\la_i}$ is a direct sum of finite dimensional
irreducible representations of $\g$.
We denote by $\V(x_1,\dots,x_N)$ corresponding affine $\hg$
representation with spectral parameters $x_i$:
\be
\label{Vdecompos0}
\V(x_1,\dots,x_N)=\bigoplus_{i=1}^N\V_{\la_i}(x_i)
\ee
We consider the intertwining operator
\beas
 && H(x_1,\dots,x_N):\\
&& \quad \V(x_1,\dots,x_N)\ot \V(x_1,\dots,x_N)\rightarrow
\V(x_1,\dots,x_N)\ot \V(x_1,\dots,x_N),
\\
 && \quad {[}H(x_1,\dots,x_N),\D(a){]}=0, \qquad \forall a\in\hg
\enas
If $\V=\V_\la$
consists of one irreducible component then $H$ is a multiple of identity,
because the tensor product is irreducible in this case.
To carry out the general case let
us gather all equivalent multiplets together:
\footnote{The $\hg$-equivalence of $\V_{\la_i}(x_i)$ requires
that the spectral parameters
$x_i$ and highest weights $\la_i$ are the same. }
\be
\label{Vdecompos}
 \V(x_1,\dots,x_N)=\bigoplus_{i=1}^M \Na_{\la_i}\hot \V_{\la_i}(x_i),
\ee
where all $\V_{\la_i}(x_i)$ are $M$ nonequivalent irreps and
$\Na_{\la_i}\simeq \ab{C}^{N_i}$ have a dimension equal to
the multiplicity of $\V_{\la_i}(x_i)$
in $\V(x_1,\dots,x_N)$.
Note that $\sum_{i=1}^M N_i=N$.
By the hat over the tensor product we mean that $\hg$ doesn't act on
$\Na_{\la_i}\hot \V_{\la_i}(x_i)$ by means of $\D$ but acts as $\id\ot g$.

So, we have:
\beas
&& \V(x_1,\dots,x_M)\ot \V(x_1,\dots,x_M)\nonumber\\
&& \quad        =(\oplus_{i=1}^M\Na_{\la_i}\hot \ \V_{\la_i}(x_i))\bigotimes
        (\oplus_{i=1}^M\Na_{\la_i}\hot \V_{\la_i}(x_i))  \\
&&\quad =
        \bigoplus_{i,j=1}^M\Na_{\la_i}\hot \Na_{\la_j}\hot
        \left( \V_{\la_i}(x_i)\ot \V_{\la_j}(x_j)\right)\nonumber
\enas
Now, $ \V_{\la_i}(x_i)\ot \V_{\la_j}(x_j)$ is equivalent only to itself and to
$ \V_{\la_j}(x_j)\ot \V_{\la_i}(x_i)$ (for $i\ne j$) by applying
the intertwining operator
$\R(x_i/x_j)=PR(x_i/x_j)$,
where $P$ is tensor product permutation: $P(v_1\ot v_2)=v_2\ot v_1$.
So,
the operator $H(x_1,\dots,x_M)$, commuting with
$\hg$ on $\V(x_1,\dots,x_M)\ot \V(x_1,\dots,x_M)$
has the following form:
\be\label{Hact}
H|_{\Na_{\la_i}\hot \Na_{\la_j}\hot \V_{\la_i}\ot \V_{\la_j}}=
        A_{ij}\hot \id_{\V_{\la_i}\ot \V_{\la_j} } +
        B_{ij}\hot \R_{\V_{\la_i}\ot \V_{\la_j} }(x_i/x_j)
\ee
where $A_{ij}$ and $B_{ij}$ are any operators on $\Na_{\la_i}\hot
\Na_{\la_j}$.
Note that in one special case the formula (\ref{Hact}) simplifies.
Consider the situation if there is only one nontrivial representation
$\V_\la$, i.e. $\la\ne 0$, in the decomposition
(\ref{Vdecompos}) of $\V(x_1,\dots,x_N)$.  As
$R_{\V_\la \ot \V_0}=R_{\V_0\ot \V_\la}={\rm id}$,
the intertwining operator (\ref{Hact}) doesn't contain  spectral
parameters $x_i$ and deformation parameter $q$. So, $H$ is invariant
under the action of quantum group for all values of deformation parameter.
Of course, this is much larger symmetry and it can be written in another form
without using of $q$.

To write down the action (\ref{Hact}) more explicitly we
introduce
the projection operators
\be
\label{proj}
X^a_b=|a\rangle\langle b|,
\ee
 where the  vectors
$|a\rangle$ span the space $\V$. In accordance with the decomposition
\re{Vdecompos} we
use the double index $a=\gr{n_i}{a_i}$, $i=1,\dots,M$
where the first
index  $n_i=1,\dots,N_i$ characterizes the multiplicity of
$\V_{\la_i}$ and the second one $a_i=1,\dots,\dim{\V_{\la_i}}$
is the vector index of $\V_{\la_i}$. Then the formula (\ref{Hact}) can be
represented as
\bea
\label{HactMat}
H(A,B)=
\sum_{i,j=1}^M
\left(
\sum_{n_i,n_j,m_i,m_j}
A_{ij}\mbox{}_{n_in_j}^{m_im_j}
\sum_{a_i,a_j}
X^\gr{n_i}{a_i}_\gr{m_i}{a_i} \ot X^\gr{n_j}{a_j}_\gr{m_j}{a_j}\right.
\\
\left.
+ \sum_{n_i,n_j,m_i,m_j}
B_{ij}\mbox{}_{n_in_j}^{m_im_j}
\sum_{a_i,a_j,a_i^\prime,a_j^\prime}
R_{ij}\mbox{}^{a_i a_j}_{a_i^\prime a_j^\prime }(x_i/x_j)
 X^\gr{n_j}{a_j^\prime}_\gr{m_i}{a_i}
\ot X^\gr{n_i}{a_i^\prime }_\gr{m_j}{a_j}
\right) ,
\nn
\ena
where $B_{ii}=0$ and we use the matrix form of $R$-operator:
\[
R_{\V_{\la_i}\ot \V_{\la_j} }(x_i/x_j) |a_i\rangle\ot|a_j\rangle =
\sum_{a_i^\prime,a_j^\prime}
R_{ij}\mbox{}^{a_i a_j}_{a_i^\prime a_j^\prime }(x_i/x_j)
|a_i^\prime\rangle\ot|a_j^\prime\rangle .
\]

Let us consider some particular cases of this general construction.

(i) Let $\V(x)=\V(x,x)=\V_\la(x)\oplus \V_\la(x)$.
	The second term in (\ref{HactMat}) is absent in this case and $H$ has
	factorized form:
\bea
\label{12}
H=A\hot \id_{\V_\la\ot \V_\la},
     &  A=A_{n_1n_2}^{m_1m_2}
\ena
where $n_1,n_2,m_1,m_2=1,2$ are indexes, which mention each $\V_\la$.
In the matrix form (\ref{12}) can be written as
\be
\label{12Mat}
H=
\sum_{n_i,n_j,m_i,m_j} A_{n_in_j}^{m_im_j}
\sum_{a_i,a_j}
X^\gr{n_i}{a_i}_\gr{m_i}{a_i} \ot X^\gr{n_j}{a_j}_\gr{m_j}{a_j}
\ee

(ii)
Let now $\V(x_1,x_2)=\V_{\la_1}(x_1)\oplus \V_{\la_2}(x_2)$ ($\V_{\la_i}(x_i)$
 are mutually
nonequivalent). Then $H$ acquires the following block-diagonal form
\be
\label{R}
H(x_1,x_2)= \left(
	\begin{array}{cccc}
		 a\cdot\id & 0 & 0 & 0 \\
        0 & c\cdot\id & d\cdot R_{12}(x_1/x_2)& 0\\
                0 & e\cdot R_{21}(x_2/x_1) & f\cdot\id & 0\\
	 0 & 0 & 0 & g\cdot\id
     \end{array}
\right)
\ee
Here to be short we used the notation
\[
R_{12}(x_1/x_2)=R_{\V_{\la_1}\ot \V_{\la_2} }(x_1/x_2), \qquad
R_{21}(x_2/x_1)=R_{\V_{\la_2}\ot \V_{\la_1} }(x_2/x_1).
\]

(iii)
Let us choose $g=sl(2)$ and $\V=\V_s\op \V_0$, where $\V_s$ is
$2s+1$-dimensional
spin-$s$ representation of
$U_qsl_2$ and $\V_0$ is the trivial one dimensional representation.
This case was considered in \cite{RA}.

Following \cite{RA} from the operator $H$ the following Hamiltonian
acting on $\W=\V^{\ot L}$ can be constructed:
\footnote{Here and in the following we omit the dependence on $x_i$}
\be
\H=\sum_{i=1}^{L-1}H_{ii+1}
\label{genhamil}
\ee
Here and in the following for the operator $X=\sum_l x_l\ot y_l$ on
$\V\ot \V$ we denote by $X_{ij}$ its action on $\W$ defined by
\be
\label{ij}
 X_{ij}=\sum_l \id\ot\dots \ot \id\ot
\underbrace{x_l}_{i}\ot \id\dots\ot\id\ot \underbrace{y_l}_{j}\ot\id\ot
	\dots\ot \id
\ee
By the construction, $\H$ is quantum group invariant:
\bea
{[}\H,\D^{L-1}(g){]}=0, & \forall g\in \hg
\nn
\ena

Let us define the projection operator $\Pro_i$ on $\V$ for each class of
equivalent irreps $(\la_i,x_i)$, $i=1,\dots,M$:
\bea
\Pro_iv_j=\delta_{ij}v_j, & \forall v_j\in \V_{\la_j}(x_j) \nn\\
\sum_{i=1}^M \Pro_i=\id, & \Pro_i^2=\Pro_i\nn
\ena
Their action on the tensor product $\W$ is given by
\[
\bar{\Pro}_i=\sum_{l=1}^L
\id\ot\ldots\id\ot\underbrace{\Pro_i}_l\ot\id\ldots\ot\id
\]
It is easy to see that these projections commute with Hamiltonian ${\cal H}$
and quantum group $\hg$:
\bea
\label{com}
[\bar{\Pro}_i,{\cal H}]=0, & [\bar{\Pro}_i,\hg]=0
\ena
Denote by
$\W_{p_1\dots p_M}$
the subspace of $\W$ with values $p_i$
of $\bar{\Pro}_i$ on it. Then we have the decomposition
\be
\label{wdecompos}
\W=\bigoplus_{\stackrel{p_1,\dots,p_M}{p_1+\dots+p_M=L}}\W_{p_1\dots p_M}
\ee

Let $\V^0$ is the linear space, spanned by the highest
weight vectors of $\g$-module $\V$:
$$
\V^0:=\op_{i=1}^N v_{\la_i}^0,
$$
where  $v^0_{\la_i}\in
\V_{\la_i}$ is highest weight vector \re{hww}.
We define also $\W^0:=\V^{0\ \ot L}$.
The space
$\W^0$ is $\H$-invariant. This
follows from (\ref{Hact}).
For general $q$ the action of $\hg$ on $\W^0$ generate whole
space $\W$.
Indeed, the $\hg$-action on each state of type
$v^0_{\la_{i_1}}\ot\ldots\ot v^0_{\la_{i_L}}$ generates the whole space
$\V_{\la_{i_1}}\ot\ldots\ot \V_{\la_{i_L}}$, because the tensor product of
finite dimensional irreducible
representations of affine quantum group is irreducible \cite{ChP}.

Consider now the subspace $\W^0_{p_1\dots p_M}=\W^0 \cap \W_{p_1\dots p_M}$.
We have the decomposition, which is inherited from \re{wdecompos}
\be
\label{w0decompos}
\W^0=\bigoplus_{\stackrel{p_1,\dots,p_M}{p_1+\dots+p_M=L}}\W^0_{p_1\dots p_M}
\ee
Note that
$$
d_{p_1\dots p_M}:=\dim \W^0_{p_1\dots p_M}  =
\left(\begin{array}{c} L \\ p_1\ldots p_M
\end{array}\right)N_1^{p_1}\ldots N_M^{p_M}
$$

Let us define by $H_0$ and ${\cal H}_0$ the restrictions of $H$ and
${\cal H}$ on $\V^0\ot \V^0$ and
$\W^0$ correspondingly: ${ H}_0:={ H}|_{\V^0\ot \V^0}$,
${\cal H}_0:={\cal H}|_{\W^0}$.
 It follows from \re{com} that  Hamiltonians ${\cal H}$ and ${\cal H}_0$ have
block diagonal form with respect to the decompositions (\ref{wdecompos}) and
(\ref{w0decompos}) correspondingly.
Every Hamiltonian eigenvector
$w^0_{\alpha_{p_1\dots p_M}}\in
\W^0_{p_1\dots p_M}$ with energy value $E_{\alpha_{p_1\dots p_M}}$
gives rise to an irreducible $\hg$-multiplet
$\W_{\alpha_{p_1\dots p_M}}$
of dimension
\be
\label{lev}
\dim \W_{\alpha_{p_1\dots p_M}}=
                        \prod_{k=1}^M(\dim \V_{\la_k})^{p_k}
\ee
On $\W_{\alpha_{p_1\dots p_M}}$
 the Hamiltonian $\H$ is diagonal with eigenvalue
$E_{\alpha_{p_1\dots p_M}}$.
In
particular case when all $\V_{\la_i}$ are equivalent, the degeneracy levels
are  the same for all $E_{\alpha_{p_1\dots p_M}}$ and are equal to
$(\dim \V_\la)^L$.
Note that
\beas
\dim \W &=&\sum_{\stackrel{p_1\dots p_M}{p_1+\dots+
	p_M=L}}\sum_{\alpha_{p_1\dots p_M}=1}^{d_{p_1\dots p_M}}
        \dim \W_{\alpha_{p_1\dots p_M}}\\
       &=& \sum_{\stackrel{p_1\dots p_M}{p_1+\dots+ p_M=L}}
	\left(\begin{array}{c} L \\ p_1\ldots p_M \end{array}\right)
	\prod_{k=1}^M N_k^{p_k}
        \prod_{k=1}^M (\dim \V_{\la_k})^{p_k} \\
	&=&
        \left(\sum_{k=1}^M N_k\dim \V_{\la_k}\right)^L
\enas
according to the decomposition
\[
\W=\bigoplus_{\stackrel{p_1,\dots,p_M}{p_1+\dots+p_M=L}}
	\bigoplus_{\alpha_{p_1\dots p_M}=1}^{d_{p_1\dots p_M}}
        \W_{\alpha_{p_1\dots p_M}}
\]


Now, we suggest that one can obtain the energy spectrum
$E_{\alpha_{p_1\dots p_M}}$ for ${\cal H}_0$.
Then for the statistical
sum we have
\be
\label{ZH0}
Z_{{\cal H}_0}(\beta)=\sum_{\stackrel{p_1,\dots,p_M}{p_1+\dots+p_M=L}}
\sum_{\alpha_{p_1\dots p_M}=1}^{d_{p_1\dots p_M}}
\exp(\beta E_{\alpha_{p_1\dots p_M}})
\ee
Then the statistical sum of ${\cal H}$ has the following form:
\be
\label{ZH}
Z_{{\cal H}}(\beta)=\sum_{\stackrel{p_1,\dots,p_M}{p_1+\dots+p_M=L}}
\prod_{k=1}^M(\dim \V_{\la_k})^{p_k}
\sum_{\alpha_{p_1\dots p_M}=1}^{d_{p_1\dots p_M}}
\exp(\beta E_{\alpha_{p_1\dots p_M}})
\ee
So, if the underlying Hamiltonian ${\cal H}_0$ is integrable and
its eigenvectors and eigenvalues can be obtained, then
this problem is solved for ${\cal H}$ also.
Performing the quantum group on all
eigenvectors of an energy level of $\H_0$ one can obtain the whole
eigenspace ${\cal H}$ for
this level.

Consider now some particular cases of our general construction.
If we choose two equivalent representations (the first
example  above),
then $\dim \V^0=2$ and there is one term in decomposition (\ref{w0decompos}).
$H_0$ now is the most general action on $\V^0\ot\V^0$. As a particular case,
the $XYZ$ Hamiltonian in the magnetic field can be obtained. The
correspondence between the spectrum of $\H$ and $\H_0$
is the simplest in this case because the additional degeneracy of all
energy levels
is the same and for the statistical sums \re{ZH0}, \re{ZH} we have
$$
Z_{\H}(\beta)=(\dim \V_\la)^L Z_{\H_0}(\beta)=
(\dim \V_\la)^L Z_{XXZ}(\beta)
$$

Consider now the second example of $H$-operator above.
Let us choose for parameters in (\ref{R})
\bea
\label{ex2}
& a=g=e=d=1 \qquad c=f=0 &
\ena
One can renormalize $R$-matrices in (\ref{R}) to satisfy the
unitarity condition
\[
R_{12}(z)R_{21}(z^{-1})=\id
\].
Together with (\ref{ex2}) this leads to
\be
\label{inv}
H(x_1,x_2)^2=\id
\ee
The
restriction of $\H$ on $\W^0$ coincides with the $XXX$ Heisenberg spin
chain
\[
\H_0=H_{XXX}=\sum_i P_{ii+1}={1\over 2}
	\sum_i \left({1+\vec{\sigma}_i \vec{\sigma}_{i+1}}\right)
\]
The space $\W^0_{p_1p_2}$, corresponds to all states
with $s_z=p_1/2$ value of spin projection
$
S^z={1/ 2}\sum_i \sigma^z_i
$.
If we return to $\H$, the  energy level degeneracy of each eigenstate with
the spin projection $s_z$ is multiplied by
$(\dim \V_{\la_1})^{2s_z}
(\dim \V_{\la_2})^{L-2s_z}$.

\setcounter{equation}{0}

\section{Generalization to long-range interaction spin chains}

Let us consider now the generalization of the construction above for the case
of long-range interacting Hamiltonians.

Recall the Haldane-Shastry
spin chain is given by \cite{Haldane88,Shastry88,Inoz90}
\be
\label{hal}
\H^{HS}_0=\sum_{i<j} {1 \over d_{i-j}^2} P_{ij},
\ee
Here the spins  take values in the fundamental representation
of $sl_n$ and $P_{ij}$ permutes the spins at $i$-th and $j$-th
positions.
It is well known that the Hamiltonian (\ref{hal}) is integrable if
$d_i$  has one of the following  values
\[
d_j=\left\{
	\begin{array}{ll}
	 j, & {\rm rational\ case}\\
	(1/\alpha)\sinh(\alpha j),\ \alpha\in {\bf{\rm R}}, &
		{\rm hyperbolic\ case}   \\
	(L/\pi)\sin(\pi j/L), & {\rm trigonometric\ case}
	\end{array}
\right.
\]
The trigonometric model is defined on periodic chain and the sum in
(\ref{hal})  is performed over
$1\le i,j\le L$.   Rational and
hyperbolic models are defined on infinite chain.

For simplicity we consider here  $sl_2$ Haldane-Shastry model only.
This means that the space $\V^0$ should be two dimensional,
so we should have two terms in the decomposition \re{Vdecompos0}.

Now our goal is to construct a long-range interaction model, which
on the highest weight space $\W^0$ coincides with \re{hal}.
So, we take $\V=\V_{\la_1}\op \V_{\la_2}$.  This is the second
example considered in sec.3. To achieve an permutation action on
the subspace $\V^0\ot \V^0$ let us impose the
conditions (\ref{ex2}) on parameters of $H$-operator \re{R}.
One can try to generalize the Hamiltonian (\ref{hal}) in
this way by
\[
\H^{HS}=\sum_{i<j} {1 \over d_{i-j}^2} H_{ij}.
\]
But it is easy to see that then $\H^{HS}$ isn't invariant with respect to
quantum group $\hg$. This is because the equation
\[
\R_{ij}(x_1,x_2)\D^{L-1}(g)=\D^{L-1}(g)\R_{ij}(x_1,x_2), \quad g\in \hg
\]
is valid only for $i=j\pm1$.

To overcome this difficulty let us use instead of $H_{ij}$
the operator
\footnote{To be short,  we omit  $x_i,x_j$-dependence
of operators $H,G,F$-operators in this section.}
\bea
\label{F}
& F_{[ij]}=G_{[ij]}H_{j-1j}G_{[ij]}^{-1},
\quad {\rm where} \quad
G_{[ij]}=H_{ii+1}
H_{i+1i+2}\ldots H_{j-2j-1} &
\ena
We remark that $F_{[ij]}$ and $G_{[ij]}$ act nontrivially on all indexes
$i,i+1,\dots,j$. So we include them into bracket do not
confuse with the definition (\ref{ij}).
The 'nonlocal' term like $F_{[ij]}$ appeared earlier
in the construction of quantum group invariant periodic spin chains
as a boundary term
\cite{Martin,GPPR,Kar}.

The operators $H_{ii+1}$ satisfy the permutation group relations
\re{inv} and
\bea
\label{braid}
 H_{i-1i}H_{ii+1}H_{i-1i}= H_{ii+1}H_{i-1i}H_{ii+1}
\ena
These relations obey on  $\W^0$, because the restriction
of $H_{ij}$ on it gives rise to permutations $P_{ij}$
\[
H_{ij}|_{\W_0}=P_{ij}
\]
On the whole space $\W$ the equations \re{braid} are continued using the affine
quantum group symmetry of $H_{i,i+1}$.
In contrast to the
standard realization by $P_{ij}$,  the relation
\be
\label{perm}
P_{i-1i}P_{ii+1}P_{i-1i}=P_{i-1i+1}
\ee
isn't fulfilled.
It is easy to see
from (\ref{F}) and \re{perm} that
\[
F_{[ij]}|_{\W^0}=P_{ij}
\]
So, the spin chain defined by
\be
\label{genhal}
	\H^{HS}=\sum_{i<j} {1 \over d_{i-j}^2} F_{[ij]},
\ee
is quantum group invariant and its restriction on the space
$\W^0$ it coincides with the Haldane-Shastry spin chain (\ref{hal}).

In some cases nonlocal expression for $F_{[ij]}$ \re{F} becomes
depending on sites $i$ and $j$ only.
For example, if $\V_{\la_1}$ is
trivial ($\V_{\la_1}=\V_0$, $\V_{\la_2}=\V_\la$), then
$R_{12}(x_1/x_2)=R_{21}(x_2/x_1)=\id$ and $H$-matrix \re{R}
has the following block-diagonal form on $\V\ot \V$ with respect to
the decomposition $\V=\V_0\op\V_\la$
\be
\label{RHS}
H= \left(
	\begin{array}{cccc}
                \id & 0 & 0 & 0 \\
        0 &0 & \id & 0\\
                0 &\id & 0 & 0\\
         0 & 0 & 0 & \id
     \end{array}
\right)
\ee
So, the block-diagonal form of $H$ is permutation and,
hence, obeys the relations \re{perm}.
Then the constructing blocks $F_{[ij]}$
act nontrivially  on two sites $i,j$ only, where they are the
block-diagonal permutations \re{RHS}.
Now all the terms in \re{genhal} are pairwise interactions between
two different sites and it can be written in explicit form
\[
        \H^{HS}=\sum_{i<j}
 {1 \over d_{i-j}^2} \left[
        \sum_{a=1}^{\dim \V_\la}
        \left(        X_i\mbox{}_0^a X_j\mbox{}_a^0 +
        X_j\mbox{}_a^0 X_i\mbox{}_0^a
        \right)+
         X_i\mbox{}_0^0 X_j\mbox{}_0^0 +
        \sum_{a,a^\prime=1}^{\dim \V_\la}
        X_i\mbox{}_a^{a} X_j\mbox{}_{a^\prime}^{a^\prime}
        \right]
\]

The energy levels of $\H^{HS}$ coincide with the levels of
(\ref{hal}). The degeneracy degree with respect to the later
is defined by (\ref{lev}). The relations (\ref{ZH0}), (\ref{ZH}) between
the statistical sums of $H_0^{HS}$ and $H^{HS}$  remain the same.



\setcounter{equation}{0}
\section{Fermionic representations}
In this section we consider the representations of some Hamiltonians
derived in previous sections in terms of $c_{i,\sigma}^\pm$, which are the
annihilation-creation operators of spin $\sigma=\uparrow,\downarrow$
fermion at site $i$.

Below we are dealing with the quantum group $\hsl$ only.
We consider the simplest cases, where
in the decomposition of $V$ there is only one representation of
spin $>0$. Then $R$-matrices, appearing in Hamiltonian are
permutations and the Hamiltonian doesn't depend on $q$. So, it
will be $\hsl$-invariant for all values of deformation parameter
$q$.

We identify the space  $\V$ with the space of spin-$\sigma$ ($\sigma=\up,\dow$)
fermionic wavefunctions.
Let us denote by $|0\rangle$ the fermionic vacuum:
$c_\sigma|0\rangle=0$. Also we use
$|\sigma\rangle=c_\sigma^+|0\rangle$,
$|\dow\up\rangle=-|\up\dow\rangle=c_\dow^+c_\up^+|0\rangle$.

To construct the fermionic representations of Hamiltonians,
introduced in the previous sections, it is convenient to
use the fermionic representation of  projection operators \re{proj}, introduced
earlier.
 Here we denote $a,b=0,\pm 1, 2$, where $0$ means the empty site,
$\pm 1$ mean
the sites $|\up\rangle$ and $|\dow\rangle$ correspondingly and
$2$ denotes the site $|\dow\up\rangle$ with double occupation.
Creation and annihilation operators can be expressed through the
projection operators as follows
\bea
\label{proj1}
c_{\sigma}^+ = X^\sigma_0-\sigma X^2_{-\sigma} &
c_{\sigma} = X^0_\sigma-\sigma X^{-\sigma}_2 &
n_{\sigma} = X^\sigma_\sigma+ X^2_2 \qquad \sigma=\pm1
\ena
and
vice versa
\bea
\label{proj2}
 X^\sigma_0=(1-n_{-\sigma})c_\sigma^+ &
X^2_\sigma=n_{\sigma}c_{-\sigma}^+ & X^0_2=c_\up c_\dow \nn\\
 X^\sigma_\sigma=(1-n_{-\sigma})n_\sigma &
X^0_0=(1-n_{\dow})(1-n_\up)=n^h & X^2_2=n_\up n_\dow=d
\ena
Other formulae are obtained using $(X^a_b)^+=X^b_a$. Here we introduced
the local Hubbard interaction operator $d$, which counts the
double occupation of site and the hole number operator $n^h$.

\subsection{$t-J$ model for zero spin-spin coupling and
Hubbard model in infinite repulsion limit}

Consider the simplest case, when the space of states decomposes
into direct sum of fundamental spin-$1/2$ and trivial spin-$0$
multiplets $\V=\V_0\op \V_{1/2}$ \cite{RA}.
We associate $\V_0$ with empty state $\br{0}$ and $\V_{1/2}$
with $\br{\up}, \ \br{\dow}$ one electron states.
Using \re{R} for the present case, we obtain
\beas
{\cal H}(t,W_1,W_2)=\sum_{i=1}^{L-1}
\left[
-t\sum_{\sigma,\sigma^\prime=\pm1}
(X_i\mbox{}^\sigma_0 X_{i+1}\mbox{}^0_\sigma+
X_{i+1}\mbox{}^\sigma_0 X_{i}\mbox{}^0_\sigma)
+ W_1 X_{i}\mbox{}^0_0 X_{i+1}\mbox{}^0_0 \right.\nn
\\
\left.+W_2\sum_{\sigma,\sigma^\prime=\pm1}  X_i\mbox{}^\sigma_\sigma
X_{i+1}\mbox{}^{\sigma^\prime}_{\sigma^\prime}
\right]
\enas
After substitution of fermionic representation \re{proj2} of projection
operators
the Hamiltonian above (by canceling out nonessential in the
thermodynamic limit boundary and
constant terms) transforms into
\bea
\label{RA1}
{\cal H}_{t-J}(t,W,\mu) = \sum^{L-1}_{i=1} \left[ -t \sum_{\sigma=\pm1}
(1-n_{i,-\sigma})( c^{+}_{i,\sigma} c_{i+1,\sigma} +
c^{+}_{i+1,\sigma}c_{i,\sigma}  ) (1-n_{i+1,-\sigma})
\right.\nn
\\
\left. +W  n_{i}n_{i+1} \right] +  \mu \sum_{i=1}^L n_{i},
\ena
where $W=W_1+W_2$ and $\mu=-2W_1$. As ${\cal H}(t,W,\mu)$
conserves the particle number, it remains integrable without
last constrain on $\mu$.
This Hamiltonian forbids double occupied sites and coincides with one of
$t-J$ model for vanishing spin-spin coupling $J=0$.
Integrability of this model and its correspondence to
$XXZ$ spin chain is well known \cite{Schlot}.
For vanishing density-density interaction $V=0$ \re{RA1} reduces
to Hubbard model in infinite repulsion limit.


\subsection{Extended Hubbard with doping, pair hopping and
hole density-density interaction }

Let consider now the case $\V=\V_0\oplus \V_1$. This corresponds to
$R_{12}(x)=R_{21}(x)=\id$ in (\ref{R}). Note that in this case
$H$ doesn't depend on $x_i$ and $q$ and we choose the coefficients there
as follows
\be
\label{HXXZ}
H= \left(
	\begin{array}{cccc}
		 W_1\cdot\id & 0 & 0 & 0 \\
        0 & 0 & -t\cdot \id& 0\\
                0 & -t\cdot \id & 0 & 0\\
	 0 & 0 & 0 & W_2\cdot\id
     \end{array}
\right)
\ee
We identify the empty state $|0\rangle$ with $\V_0$ and the space spanned
by $|\up\rangle$, $|\dow\up\rangle$, $|\dow\rangle$ with $\V_1$.
The general Hamiltonian \re{genhamil}  in our case has the form
\bea
\label{RA01}
{\cal H}(t,W_1,W_2)=\sum_{i=1}^{L-1}
\left[
-t\sum_{\sigma=\pm1}
(X_i\mbox{}^\sigma_0 X_{i+1}\mbox{}^0_\sigma+
X_{i+1}\mbox{}^\sigma_0 X_{i}\mbox{}^0_\sigma)
-t( X_i\mbox{}^2_0 X_{i+1}\mbox{}^0_2
\right.\nn\\
+X_{i+1}\mbox{}^2_0 X_{i}\mbox{}^0_2)
+W_1 X_{i}\mbox{}^0_0 X_{i+1}\mbox{}^0_0
\left.
+W_2\sum_{\sigma,\sigma^\prime=\pm1}
(X_i\mbox{}^\sigma_\sigma +X_i\mbox{}^2_2)
(X_{i+1}\mbox{}^{\sigma^\prime}_{\sigma^\prime}+X_{i+1}\mbox{}^2_2)
\right]
\ena
After substituting \re{proj2} into \re{RA01} one obtain
\bea
\label{H01}
{\cal H}(t,W_1,W_2)=\sum^{L-1}_{i=1} \left[ -t \cdot \sum_{\sigma} ( 1-n_{i,-\sigma} )
( c^{+}_{i,\sigma}c_{i+1,\sigma} + c^{+}_{i+1,\sigma}c_{i,\sigma} )
( 1-n_{i+1,\sigma} ) \right. \nn
\\
-   t ( c^{+}_{i,\uparrow}c^{+}_{i,\downarrow}c_{i+1,\downarrow}
c_{i+1,\uparrow} + c^{+}_{i+1,\uparrow}c^{+}_{i+1,\downarrow}c_{i,\downarrow}
c_{i,\uparrow} )  +     W_1 n_i^h n_{i+1}^h
\\
\left. +W_2  (n_{i}-d_i)(n_{i+1}-d_{i+1})\right]
-  2W_2\sum_{i=1}^L  n_i
   +W_2(n_1^h+n_L^h) + W_2L\nn
\ena
The quantum group generators (\ref{QG}) (after some
renormalization)  can be written in terms of
fermionic operators also:
\bea
\label{QGFerm}
e_1=[2]_q^{-1/2} (n_\up c^+_\dow+n_\dow c_\up)& e_0=f_1 \nn\\
f_1=[2]_q^{-1/2} (n_\up c_\dow+n_\dow c^+_\up)& f_0=e_1\\
h_1=2(n_\up-n_\dow)& h_0=-h_1\nn
\ena
The first term in (\ref{H01}) is known as "doping" term, the second one
is pair hopping term. Using $n^h=1-n+d$ and omitting unessential boundary
and constant terms in  (\ref{H01}) one obtain
\bea
\label{H01hole1}
{\cal H}(t,W,U,\mu)=\sum^{L-1}_{i=1} \left[ -t  \sum_{\sigma} (
1-n_{i,-\sigma} )
( c^{+}_{i,\sigma}c_{i+1,\sigma} + c^{+}_{i+1,\sigma}c_{i,\sigma} )
( 1-n_{i+1,\sigma} ) \right.
\nn\\
 - \left. t ( c^{+}_{i,\uparrow}c^{+}_{i,\downarrow}c_{i+1,\downarrow}
c_{i+1,\uparrow} + c^{+}_{i+1,\uparrow}c^{+}_{i+1,\downarrow}c_{i,\downarrow}
c_{i,\uparrow} )
+ W n^h_i n^h_{i+1} \right]
\\
+ U  \sum_{i=1}^L n_{i,\uparrow}n_{i,\downarrow}
+ \mu  \sum_{i=1}^L n_{i}
\nn,
\ena
where $W=W_1+W_2$, $\mu=2W_2$, $U=-2W_2$.

In view of the consideration, carried out in sec.3, the energy levels
of obtained fermionic model coincide with the levels of
spin chain model $\H_0(t,W_1,W_2)$ on $\W^0$. It is easy to see
that its two site interaction term
\be
\label{H0XXZ}
H_0= \left(
        \begin{array}{cccc}
                 W_1 & 0 & 0 & 0 \\
        0 & 0 & -t& 0\\
                0 & -t & 0 & 0\\
         0 & 0 & 0 & W_2
     \end{array}
\right)
\ee
leads to $XXZ$ model
in external magnetic field (see \re{XXZ}, \re{DeltaW} below).

We note that the Hamiltonian \re{H01} preserves
the number of holes
${\cal N}^h=\sum_{i=1}^L n_i^h$ and the number of double occupied states
${\cal D}=\sum_{i=1}^L d_i$. So, we can choose the
parameters of corresponding terms in (\ref{H01hole1}) to be
arbitrary without loose of integrability.
Of course, the Hamiltonian (\ref{H01hole1}) is not invariant
with respect to quantum group (\ref{QGFerm}) now.

\subsection{Hubbard model with bond-charge interaction
and additional density-density and boson-boson interactions }

 Let us consider  $\V(x_a,x_b)=\V_{a}(x_a)\op\V_{b}(x_b)$
and identify
the first multiplet in this decomposition ($\V_a$) with the  space spanned by
$|\up\rangle,|\dow\rangle$ and  the second one ($\V_b$)
with the space spanned by
$|0\rangle,|\up\dow\rangle$. We consider two different commuting quantum group
actions on $\V$. The first one (${\hsl}^{(1)}$) acts on $\V_a$ as
spin-$1/2$ representation ($\V_a\sim\V_{1/2}$) and on $\V_b$ as two
spin singlets ($\V_b\sim\V_0\op\V_0$).  The second one
(${\hsl}^{(2)}$) acts vice versa: $\V_a\sim\V_0\op\V_0$ and
$\V_b\sim\V_{1/2}$.

Now by use of the form of intertwining operator $H$ \re{HactMat}
for the decomposition $\V=\V_{1/2}\op\V_0\op\V_0$ and taking
vanishing matrixes
$A_{0\frac12}$ and $A_{\frac12 0}$ ,
one  obtain
\bea
\label{H12Mat}
{\cal H}(A,B)=
\sum_{\langle i,j \rangle}\left[
\sum_{\delta^i,\delta^{i\prime},\delta^j,\delta^{j\prime}}
A_{00}\mbox{}_{\delta^i\delta^j}^{\delta^{i^\prime}\delta^{j^\prime}}
X_i\mbox{}^{\delta^i}_{\delta^{i^\prime}}
X_j\mbox{}^{\delta^j}_{\delta^{j^\prime}}+
\sum_{\sigma^i,\sigma^j}
A_{\frac12\frac12}
X_i\mbox{}^{\sigma^i}_{\sigma^i}
X_j\mbox{}^{\sigma^j}_{\sigma^j}\right.
\nn\\
\left.
+\sum_{\sigma,\delta^i,\delta^j}
\left(
B_{0\frac12}\mbox{}^{\delta^i}_{\delta^j}
X_i\mbox{}^{\sigma}_{\delta^i} X_j\mbox{}^{\delta^j}_{\sigma}+
B_{\frac12 0}\mbox{}_{\delta^i}^{\delta^j}
X_i\mbox{}_{\sigma}^{\delta^i} X_j\mbox{}_{\delta^j}^{\sigma}\right)\right]
\ena
Note that the indexes $i,j$ here denote the nearest neighbors but not
nonequivalent representations $\la_i$, as in \re{HactMat}.
So we used the upper index for them  not to mix with \re{HactMat}.
Also, we changed the
double indexes in \re{HactMat} on single indexes $\sigma=\pm1$, $\delta=0,2$ by
$
(1,a_{\frac12})\sim \sigma,\ (n_0,1)\sim \delta.
$
Let us
choose the parameters in \re{H12Mat} in the following way
\[
A_{00}\mbox{}_{\delta^i\delta^j}^{\delta^{i^\prime}\delta^{j^\prime}}=
\delta_{\delta^i}^{\delta^{i^\prime}}\delta_{\delta^j}^{\delta^{j^\prime}}W_2,
 \quad A_{\frac12\frac12}=W_1,
\quad
B_{0\frac12}\mbox{}^{\delta^i}_{\delta^j}=
B_{\frac12 0}\mbox{}^{\delta^i}_{\delta^j}=
-t\delta_{\delta^j}^{\delta^i}.
\]
After these simplifications we have
\bea
\label{sh}
{\cal H}(t,W_1,W_2)=\sum_{i=1}^{L-1}
\left[
-t\sum_{
\stackrel{\sigma,\sigma^\prime=\pm1}
{\mbox{}_{\delta,\delta^\prime=0,2}}
}
(X_i\mbox{}^\sigma_\delta X_{i+1}\mbox{}^\delta_\sigma+
X_{i+1}\mbox{}^\sigma_\delta X_{i}\mbox{}^\delta_\sigma)
\right.\nn\\
+W_1\sum_{\sigma,\sigma^\prime=\pm1}  X_i\mbox{}^\sigma_\sigma
X_{i+1}\mbox{}^{\sigma^\prime}_{\sigma^\prime}
\left. +W_2\sum_{\delta,\delta^\prime=0,2}
X_{i}\mbox{}^\delta_\delta
X_{i+1}\mbox{}^{\delta^\prime}_{\delta^\prime}
\right]
\ena
The nearest neighbor interaction of this Hamiltonian conserves
its form with respect to  index exchange $\delta\leftrightarrow\sigma$.
This means that $\H(t,W_1,W_2)$ is invariant
with respect to both quantum groups ${\hsl}^{(1)}$ and ${\hsl}^{(2)}$. So,
\re{sh} has $\hsl\ot\hsl$-symmetry. The highest weight space at each
site is formed by the two vectors $\br{\up}$ and $\br{0}$ of
$\V_a$ and $\V_b$ correspondingly.

Using the fermionic representation of $X$-operators (\ref{proj2}) we
obtain
\bea
\label{sh1}
{\cal H}(t,W_1,W_2)=\sum_{i=1}^{L-1}\left[
\sum_{\sigma=\up,\dow}
\left\{
	-t ( c^{+}_{i,\sigma}c_{i+1,\sigma}+c^{+}_{i+1,\sigma}c_{i,\sigma})
	\right.\right.\nn\\
	\left. + t (c^{+}_{i,\sigma}c_{i+1,\sigma}+c^{+}_{i+1,\sigma}c_{i,\sigma})
	(n_{i,-\sigma}+n_{i+1,-\sigma})\right\}
	\\
	+(W_1+W_2)( n_{i}n_{i+1}-2n_id_{i+1}-2n_{i+1}d_i+4d_id_{i+1} )
	\nn\\
	+\left. 2W_1(d_i+d_{i+1}) - W_1(n_i+n_{i+1}) +W_1\right]\nn
\ena
Two types of hopping are allowed here. First, fermion from single
occupied side can hope to empty site and vice versa. Second,
from double occupied site fermion can hope to single occupied site,
which contains fermion with opposite spin. Such hopping term
 appeared in \cite{Klein} after projection of Hubbard model on
 a fixed occupation number subspace.
Thus, the Hamiltonian \re{sh1} preserves the occupation
number of fermions: number
of holes ${\cal N}^h$ and number of double occupied sites ${\cal D}$.
So, corresponding terms without affecting on the integrability
an be rewritten with arbitrary coefficients. Also, note that the term
$$
2(W_1+W_2)\sum_{i=1}^{L-1}2d_id_{i+1}-n_id_{i+1}-n_{i+1}d_i
$$
can be written as hole and double occupation site interaction
and Hubbard
interaction in the following way
$$
2(W_1+W_2)\sum_{i=1}^{L-1}n^h_id_{i+1}+n^h_{i+1}d_i-d_i-d_{i+1}
$$
Thus one can generalize \re{sh1} by
\bea
\label{sh2}
{\cal H}(t,W_1,W_2,U,\mu)=\sum_{i=1}^{L-1}
\sum_{\sigma=\up,\dow}
\left[
	-t (c^{+}_{i,\sigma}c_{i+1,\sigma}+c^{+}_{i+1,\sigma}c_{i,\sigma})
	\right.
	\nn\\
	\left. + t (c^{+}_{i,\sigma}c_{i+1,\sigma}
	+c^{+}_{i+1,\sigma}c_{i,\sigma})
	(n_{i,-\sigma}+n_{i+1,-\sigma})\right]
	\\
	+W\sum_{i=1}^{L-1}( n_{i}n_{i+1}
	+2n_i^hd_{i+1}+2n_{i+1}^hd_i )
	+U\sum_{i=1}^{L}d_i + \mu\sum_{i=1}^{L}n_i
	\nn\\
	+W_1(n_1+n_L)+2W_2(d_1+d_L)\nn
\ena
The term with coefficient $W=W_1+W_2$ consists of standard fermion
density-density interaction between nearest neighbors and
interaction between empty site (hole) and nearest neighbor
site occupied by two fermions
with opposite spins. The later interaction can be considered as interaction
between two different bosonic sites $|0\rangle$,  $|\dow\up\rangle$.
The boundary terms, which appeared in \re{sh2},
do not affect on the spectra of \re{sh2} in the
thermodynamic limit and can be omitted. So, in this limit the
Hamiltonian above depends only on $W=W_1+W_2$.

The connection between Hamiltonians \re{sh1} and \re{sh2}
is given by
\be
\label{con}
{\cal H}(t,W_1,W_2,U,\mu)=
{\cal H}(t,W_1,W_2)+(U+4W_2){\cal D}+(\mu+2W_1){\cal N }-W_1(L-1)
\ee

Now we consider the restriction of \re{sh1} on the highest weight space,
which is generated by the empty sites and sites occupied by the single
$\sigma=\up$ spin.
Looking on the matrix form of $H_0(t,W_1,W_2)=A$ we recognize
 $XXZ$ spin chain.
It follows from our previous investigation that the obtained model
is exactly solvable and has the same energy levels as $XXZ$ Heisenberg
model in the external homogeneous magnetic field $\vec{B}=B\vec{z}$
along $z$ axes
\bea
\label{XXZ}
\H_0(t,W_1,W_2)&=&\H_{XXZ}(t,\Delta,B)
\nn\\
&=&-\frac{t}{2}\sum_{i=1}^{L-1} \left(
\sigma^x_i \sigma^x_{i+1}+\sigma^y_i \sigma^y_{i+1} +
\Delta\sigma^z_i \sigma^z_{i+1} +\frac B2\sigma^z_i\right),
\ena
where
\be
\label{DeltaW}
\Delta=-\frac W{2t}, \qquad
B=\frac 2t(W_1-W_2)
\ee
 All energy levels
have the same degeneracy.
Recall that  we identified the spin states $|+\rangle,|-\rangle$
of \re{XXZ} with the highest weight states $|\up\rangle$ and
$|0\rangle$ of $V$ correspondingly.

It is easy to check that the Hamiltonian \re{sh}, \re{sh1} commutes
with $\eta$-pairing operators \cite{Yang,YZ,EKS1,EKS2}:
\bea
\label{eta}
\eta=\sum_{i=1}^L c_{i,\up}c_{i,\dow} =\sum_{i=1}^L X_i\mbox{}^0_2 \qquad
\eta^+=\sum_{i=1}^L c^+_{i,\dow}c^+_{i,\up} =\sum_{i=1}^L X_i\mbox{}^2_0
\nn\\
\eta^z= {\cal N}-L=\sum_{i=1}^L (X_i\mbox{}^2_2-X_i\mbox{}^0_0),
\ena
which generate $sl_2$ algebra.
Using this fact and \re{con} one can compute the commutation rules of \re{eta}
with
generalized Hamiltonian \re{sh2}
\bea
\label{Heta}
{[} \H (t,W_1,W_2,U,\mu),\eta {]}=-(U+2\mu+4W)\eta
\nn\\
{[} \H (t,W_1,W_2,U,\mu),\eta^+{]}=(U+2\mu+4W)\eta^+
\ena
So, we have $[\H(t,W_1,W_2,U,\mu),\eta^\pm]=0$ if parameters
satisfy the condition $\mu=-U/2-2W$.

We are interested in thermodynamic behavior at $T=0$.
Denote by $E_N=E_N(t,W_1,W_2,U,\mu)$ the ground state energy of
$\H(t,W_1,W_2,U,\mu)$ in the $N$-particle sector. Also, let
${\cal E}_{N_1}={\cal E}_{N_1}(\Delta)$ is the ground state of $XXZ$ spin chain
\re{XXZ} without external magnetic field ($B=0$)
in the sector with $N_1$ upturned spins.
Recall that single occupied sites of particle space correspond to
positive spin in $XXZ$ model whereas double and empty sites correspond
to negative spin.
Thus, in the sector, which consists of states with fixed  occupation number
${\cal N}_1=N_1$, ${\cal D}=(N-N_1)/2$ the minimal energy $E_N^{N_1}$
of $\H(t,W_1,W_2,U,\mu)$ is (see \re{con})
\be
\label{ENN1}
E_N^{N_1}={\cal E}_{N_1}-\left(\frac{U}2+W\right)N_1+
\left(\frac{U}2+\mu+2W\right)N
+\frac L2(W_1-W_2)
\ee
To obtain ground state $E_N$ one should minimize \re{ENN1}
among all possible values of ${ N}_1$
\bea
\label{EN}
E_N=
\min_{0\le N_1 \le N} E_N^{N_1}=
\min_{0\le N_1 \le N} \left[
{\cal E}_{N_1}-\left(\frac{U}2+W\right)N_1\right]
\\
+\left(\frac{U}2+\mu+2W\right)N+\frac L2(W_1-W_2)\nn
\ena
Note that the Hamiltonians $\H_{XXZ}(t,\Delta,0)$ and $\H_{XXZ}(-t,-\Delta,0)$
are related by similarity transformation \cite{Gaudin}. So, the
spectrum of $\H_{XXZ}(t,\Delta,0)$ in the ferromagnetic regions
$(-t>0,\Delta\le-1)$ and $(-t<0,\Delta\ge1)$ coincides.
Hereafter we dial with the first region and choose the parametrization
$\Delta=-\cosh\gamma$, $\gamma>0$.
Note that \re{XXZ} in zero magnetic field has two $Z_2$
symmetric ferromagnetic ground states
$|\pm,vac\rangle_{XXZ}=|\pm,\pm,\dots,\pm\rangle$ with energy
${\cal E}_0={\cal E}_L=-t/2(L-1)\Delta$.

The energy levels of $XXZ$ spin chain are determined by means of
Bethe Ansatz \cite{Bethe,YY}. In the thermodynamic limit $L\to\infty$
there
it has string type solutions. It was shown in \cite{AKS} that
the minimal energy in the sector with $N_1$ upturned spins is
\be
\label{N1st}
{\cal E}_{N_1}=-2t\sinh\gamma\frac{\sinh N_1\gamma}{\cosh N_1\gamma+1}
+{\cal E}_0
=-2t\sinh\gamma\tanh \frac{N_1}2\gamma+{\cal E}_0
\ee
and corresponds to single string of length $N_1$.
Here the magnetization is restricted by $0\le N_1/L\le1/2$, i.e.
we are dealing with excitations near ferromagnetic vacuum $|-,vac\rangle$.
Due to $Z_2$ symmetry we have ${\cal E}_{N_1}={\cal E}_{L-N_1}$ and
the excitations near $|+,vac\rangle$ ($1/2\le N_1/L\le 1$) can be considered
in a similar way
by exchanging $N_1\to L-N_1$ in \re{N1st}.

For infinite number of turned spins
$\lim_{L\to\infty} N_1/L=0,L$ the minimal energy \re{N1st} is
${\cal E}_\infty=-2t\sinh\gamma+{\cal E}_0$. So, all ground states
for fixed magnetizations, which differ from both ferromagnetic ground
states by infinite number of turned spins, have the same energy.

It is well known that $XXZ$ spin chain for $|\Delta|>1$ is massive.
The energy gap between ground state and elementary excitations in the limit
$N_1\to\infty$ is $\Delta E=-2t(\cosh\gamma-1)$.

Substituting \re{N1st} into \re{EN} one obtain the value of $N_1$,
which minimize $E_N^{N_1}$ in \re{EN}
\be
\label{N1min}
N_1^{min}=\left\{
	\begin{array}{ll}
	0 & \left( U/2+W\le0\right)\quad {\rm or} \quad
	 \left(0< U/2+W\le\gamma/2, \ N\le N_c\right),\\
	N & \left( U/2+W>0\right)\quad {\rm or} \quad
	\left(0< U/2+W\le\gamma/2, \ N\ge N_c\right),
	\end{array}
	\right.
\ee
where $N_c$ is solution of
\[
\tanh\frac {N_c}2\gamma=\left(\frac U2+W\right)N_c, \qquad \frac U2-W>0,
\quad N_c>0
\]
Consider the  regions
$(U/2+W\le0,\ W\le2t)$
and $(0<U/2+W\le\gamma/2,\ N\le N_c,\ W\le2t)$ in \re{N1min},
where there are no
single occupied states. The ground state of Hamiltonian
in the $N$-particle sector ($N$ is even) is any state with ${\cal D}=N/2$
double occupied sites and $L-N/2$ empty sites.
Let us write down all these states in the form
\be
\label{psi}
\psi^N_P=(\eta^+_P)^{N/2}\vac, \qquad \vac=|0,0,\dots,0\rangle
\ee
Here $\eta_P^\pm$ are generalized $P$-momentum $\eta$-pairing operators
\beas
\eta_P=\sum_{j=1}^L e^{iPj}c_{j,\up}c_{j,\dow} \qquad
\eta^+_P=\sum_{j=1}^L e^{-iPj}c^+_{j,\dow}c^+_{j,\up}
\enas
Note that $\eta_0^\pm=\eta^\pm$.

All the states $\psi^N_P$ obey ODLRO, i.e.
\be
\label{ODLRO}
\lim_{|i-j|\to\infty}\frac{\langle\psi_P^N|
c_{j,\dow}^+c_{j,\up}^+c_{i,\up}c_{i,\dow}
|\psi^N_P\rangle}{\langle\psi^N_P|\psi^N_P\rangle}  \ne 0
\ee
Note also that $\eta^N_P$ operators obey $\eta^N_P\vac=0$.
As a consequence the ground states $\psi_P^N$ are superconducting.

In the phase space regions
$(U/2+W>\gamma/2,\ W\le2t)$
and $(0<U/2+W\le\gamma/2,\ N\ge N_c,\ W\le2t)$ the are no double occupied
sites in the ground states of \re{sh2}. These states are linear combinations
of states with $N$ single fermions of same spin. Of course, they do not
exhibit ODLRO \re{ODLRO}.

The projection of \re{sh2} on states without double occupation (${\cal D}=0$)
gives rise to $t-J$ Hamiltonian for zero spin-spin coupling $J=0$ \re{RA1}
with additional boundary term $W_1(n_1+n_L)$.
As it was mentioned above, this model is integrable and equivalent to
Heisenberg magnet.
Note that this equivalence can be seen from \re{RA1}. If we
restrict  $\H(t,W,\mu)$ on the states with only one sort of fermion
(spin-$\up$ or spin-$\dow$), we obtain the fermion representation of $XXZ$
model
\beas
\label{F-XXZ}
{\cal H}_{XXZ}(t,W_1,W_2,\mu)=\sum_{i=1}^{L-1}
	-t (c^{+}_{i}c_{i+1}+c^{+}_{i+1}c_{i})
	+W\sum_{i=1}^{L-1} n_{i}n_{i+1}
 + \mu\sum_{i=1}^{L}n_i
\enas

For the special values of interaction potentials $W_1=-W_2=W$ the
Hamiltonian under consideration  simplifies drastically.
We have up to boundary terms
\bea
\label{sh3}
{\cal H}(t,W)=\sum_{i=1}^{L-1}
\sum_{\sigma=\up,\dow}
\left[
	-t ( c^{+}_{i,\sigma}c_{i+1,\sigma}+c^{+}_{i+1,\sigma}c_{i,\sigma})
	\right.\nn\\
	\left. + t (c^{+}_{i,\sigma}c_{i+1,\sigma}+c^{+}_{i+1,\sigma}c_{i,\sigma})
	 (n_{i,-\sigma}+n_{i+1,-\sigma})\right]
	+U\sum_{i=1}^{L}d_i + \mu\sum_{i=1}^{L}n_i
\ena
This is  Hubbard model with bond-charge interaction.
On the highest weight space it coincides with
$XY$ model in external magnetic field (or equivalently free fermion model
with chemical potential term added) ($\Delta=0$ in \re{XXZ}).
So, ${\cal H}(t,W)$
is exactly solvable and have the same energy levels as
free fermion model with chemical potential term
with degeneracy degree of each level.

The Hamiltonian \re{sh3}
${\cal D}$ was obtained and
solved in \cite{Klein,Schad}.
In our construction its integrability
is obvious, because the restriction ${\cal H}^0(t,W=0)$
of (\ref{sh3})
on the highest
weight space coincides with ordinary $XY$ spin chain
and, consequently, it is equivalent to free fermion spin chain.
It was shown in \cite{Schad} that \re{sh3} has superconducting
ground states. Note that here we didn't consider the antiferromagnetic
region, which corresponds to $\Delta=0$ case. We suggest that
in this region \re{sh2} has superconducting ground state too,
obeying $\eta$-pairing mechanism.

\setcounter{equation}{0}
\section{Aknowlegement}

This work was supported in part by INTAS grant \#840,
and by the Grant 211-5291 YPI of the German Bundesministerium fur
Forschumg und Technologie.

\end{document}